\documentclass[12pt]{article}
\usepackage{latexsym,graphicx,a4wide,epsfig} 
\usepackage[bookmarks]{hyperref}

\newcommand{\bdm}{\begin{displaymath}}
\newcommand{\edm}{\end{displaymath}}
\newcommand{\no}{\nonumber \\}

\newcommand{\be}{\begin{equation}}
\newcommand{\ee}{\end{equation}}
\newcommand{\bea}{\begin{eqnarray}}
\newcommand{\eea}{\end{eqnarray}}
\newcommand{\fs}{\; \; .}
\newcommand{\co}{\; \; ,}

\newcommand{\la}{\langle}
\newcommand{\ra}{\rangle}

\newcommand{\pp}{|\mathbf{p}|}
\newcommand{\qq}{|\mathbf{q}|}

\begin{document}
\begin{titlepage}

\begin{flushright}ZU-TH 7/01\end{flushright}
\vspace{3cm}
\begin{center}{\Large\bf Dispersion relations and soft pion theorems for $K \to \pi \pi$} 

\vspace{2 cm}
M.~B\"uchler, G.~Colangelo, J.~Kambor and F.~Orellana

\vspace{1cm}
Institute for Theoretical Physics, University of Z\"urich\\
Winterthurerstr. 190, CH-8057 Z\"urich, Switzerland\\

\vspace{1cm}
February 2001
\end{center}

\vspace{4cm}

\begin{abstract}
We propose a new method to obtain the $K \to \pi \pi$ amplitude from $K \to
\pi$ which allows one to fully account for the effects of final state
interactions. The method is based on a set of dispersion relations for the
$K \to \pi \pi$ amplitude in which the weak Hamiltonian carries
momentum. The soft pion theorem, which relates this amplitude to the $K \to
\pi$ amplitude, can be used to determine one of the two subtraction
constants -- the second constant is at present known only to leading order
in chiral perturbation theory.  We solve the dispersion relations
numerically and express the result in terms of the unknown higher order
corrections to this subtraction constant.
\end{abstract}
\end{titlepage}

\noindent {\bf 1.} 
The $K\rightarrow \pi\pi$ amplitude has been the subject of numerous
studies over the years, the main challenge being either to explain the
$\Delta I=1/2$ rule or to provide a reliable estimate of
$\varepsilon'/\varepsilon$ \cite{list,lattice}.  Among these attempts, the
lattice approach is in principle the most rigorous as the weak matrix
elements are calculated from first principles in a truly nonperturbative
way. However, the inclusion of final state interactions is problematic. The
calculation of the $K\rightarrow \pi\pi$ amplitude, e.g., proceeds by
calculating the $K\rightarrow \pi$ amplitude on the lattice and using
chiral perturbation theory (CHPT) at tree level to obtain the physical
decay amplitude \cite{Bernard}.
As is well known, the latter step induces a sizable uncertainty in the
final result, commonly estimated to be around 30\%, the typical size of
next--to--leading order corrections in chiral SU(3).  To discuss the
relation between the $K \to \pi \pi$ and the $K \to \pi$ amplitude, it is
necessary to allow the weak Hamiltonian to carry momentum. Then the former
amplitude becomes a function of the usual three Mandelstam variables $s,\;
t$ and $u$, and is identified with the physical decay amplitude at the
point $s=M_K^2$, $t=u=M_\pi^2$.  At the so--called soft--pion point (SPP),
where the momentum of one of the two pions is sent to zero ($s=u=M_\pi^2$,
$t=M_K^2$), this amplitude is related to the $K \to \pi$ amplitude, up to
${\cal O}(M_\pi^2)$ corrections. The problem is how to extrapolate the
amplitude from the SPP to the physical point. The only working
method proposed so far has been to use CHPT at tree level \cite{Bernard} --
using the one-loop relation does not solve the problem because a number of
unknown low energy constants appear \cite{BPP}.

In the present letter we set up a dispersive framework for the $K \to \pi
\pi$ amplitude. We show that by solving numerically the dispersion
relations one can do the extrapolation in a controlled manner. The
unitarity corrections due to rescattering of the pions in the final state,
and those due to $\pi K$ (virtual) rescattering in the $t$ and $u$
channels, can be accurately accounted for by solving the dispersion
relations. These effects, which also appear to one loop in CHPT, are not
the only source of possible large corrections to tree level: two
subtraction constants appear which may also suffer from large ${\cal
O}(M_K^2)$ corrections.  The soft--pion theorem provides the means to
determine one of the two subtraction constants, up to terms of order
$M_\pi^2$. The other constant (the derivative in $s$ of the amplitude at the
SPP) is unfortunately not yet determined with the same
accuracy, and at present can be estimated only with tree-level CHPT. A
better determination of this constant is the core of the problem. Once
solved the $K \to \pi \pi$ amplitude can be obtained with substantially
smaller uncertainties than at present.

Truong \cite{truong}, and more recently Pallante and Pich \cite{PP}, have
stressed the importance of final state interactions in $K \to \pi \pi$, for
the $\Delta I = 1/2$ rule and $\varepsilon'/\varepsilon$, respectively. In
estimating these effects they rely on the dispersion relation for the $K
\to \pi \pi$ amplitude with the kaon off-shell. While the method provides a
quick and simple estimate of the effect of final state interactions, it is
not trivial to promote it to a systematic and rigorous calculation. The
problems related to the formulation and the use of dispersion relations for
an off-shell amplitude are discussed in a separate note
\cite{offshellpaper} (see also \cite{suzuki}).

\vskip 0.5cm
\noindent {\bf 2.} We consider the amplitude\footnote{We discuss here only
  the $\Delta I=1/2$ amplitude -- the $\Delta I=3/2$ can be treated
  similarly.} 
\be 
\la \pi(p_1) \pi(p_2) (I=0) | {\cal H}^{1/2}_W(0)|K(q_1) \ra =: T^+(s,t,u)
\label{eq:Adef}
\ee 
described in terms of the Mandelstam variables: 
\be
s=(p_1+p_2)^2, \; \;t=(q_1-p_1)^2, \; \; u=(q_1-p_2)^2 \; \; , 
\ee 
related by $s+t+u=2M_\pi^2+M_K^2+q_2^2$, where $q_2$ is the momentum
carried by the weak Hamiltonian. From now on we set $q_2^2=0$ (but $q_2^\mu
\neq0$ in general). The physical decay amplitude is obtained by  
setting $q_2^\mu=0$ ($s=M_K^2$, $t=u=M_\pi^2$).

Since the weak Hamiltonian has the quantum numbers of the kaon, and the
pions are in an isospin zero state, the amplitude (\ref{eq:Adef}) is
analogous to the $t \leftrightarrow u$ even combination of the $KK \to \pi
\pi$ scattering amplitude (the notation is borrowed from
Ref. \cite{lang}). Like in that case, one can show that if one neglects the
imaginary parts of $D$ waves and higher in all channels, then the analytic
structure of the amplitude simplifies and it can be decomposed into a
combination of functions of a single variable (for the $K\pi$ scattering
case see \cite{anantbu}):
\bea
T^+(s,t,u)&=&M_0(s)+{1 \over 3}\left[N_0(t)+N_0(u)\right] + {2 \over 3}
\left[R_0(t)+R_0(u)\right] \no &+&{1 \over 2}\left[ \left(s-u-{M_\pi^2
\Delta \over t} \right) N_1(t)+\left(s-t-{M_\pi^2 \Delta \over u}
\right)N_1(u) \right] \co
\label{eq:T+dec}
\eea
where $\Delta = M_K^2-M_\pi^2$.
Notice that the terms proportional to $N_1$ drop out from the physical
decay amplitude:
\bea
{\cal A}(K\to \pi \pi)&=& T^+(M_K^2,M_\pi^2,M_\pi^2) = \no
&=& M_0(M_K^2)+{2\over3} \left[ N_0(M_\pi^2)+2R_0(M_\pi^2) \right] \fs
\eea

Each of the single variable functions appearing in Eq. (\ref{eq:T+dec}) is
analytic in the complex plane except for a cut starting at $4
M_\pi^2$ for $M_0$ and at $(M_K+M_\pi)^2$ for the remaining ones. 
These functions are defined to have the discontinuity on the positive real
axis identical to that of a specific partial wave: $M_0$ to the
$I=0$ $S$-wave in the $s$ channel, whereas in the $t$ channel, $N_0$ and
$N_1$ to the $I=1/2$ $S$- and $P$- wave respectively, and $R_0$ to the
$I=3/2$ $S$-wave\footnote{We disregard the imaginary part of the $I=3/2$
  $P$-wave in the $t$ channel because it is phenomenologically very
  small and vanishes in the chiral expansion up to order $p^6$}.  
Below the inelastic threshold, the elastic unitarity condition for these
functions reads
\bea
{\rm disc} M_0(s) &=&\sin \delta_0^0(s) e^{-i\delta_0^0} \left[ M_0(s)+\hat
  M_0(s) \right] \co \no
{\rm disc} N_\ell(s) &=&\sin \delta_\ell^{1/2}(s) e^{-i\delta_\ell^{1/2}}
\left[ N_\ell(s)+\hat N_\ell(s) \right] \co \no
{\rm disc} R_0(s) &=&\sin \delta_0^{3/2}(s) e^{-i\delta_0^{3/2}}
\left[ R_0(s)+\hat R_0(s) \right] \co
\label{eq:elast_unit}
\eea
where $\delta_0^0$ is the $\pi \pi$ phase shift, whereas those with
half-integer isospin are the $\pi K$ phase shifts.

The hat functions denote contributions from the other
channels coming in via angular averages (to be specified below), and are
defined as
\bea
\hat M_0(s) &=& \left[ s \Sigma_1-2 M_K^2 M_\pi^2 -
  {1\over 4}(M_K^4+3s^2) \right]\la \tilde N_1 \ra  
 + 2  s \pp \qq \la z \tilde N_1\ra \no  
&+&  4 \pp^2 \qq^2 \la z^2 \tilde N_1 \ra
+ {2 \over 3}( \la N_0 \ra + 2 \la R_0 \ra ) \co \no
\hat N_0(t)&=& \la M_0 \ra_s + y(t) \la M_1 \ra_s  
- r(t) \la z M_1 \ra_s
+{1 \over 3} \left( 4 \la R_0 \ra_u-\la N_0\ra_u \right) \no
&-& {1 \over 8} w(t) \la \tilde N_1 \ra_u
  -{1\over 4} v(t) \la z \tilde N_1 \ra_u
+{1 \over 8} r^2(t) \la z^2 \tilde N_1 \ra_u \co \no 
\hat N_1(t) &=& {2 \over r(t)} \left\{
    \la z M_0 \ra_s + y(t) \la z M_1 \ra_s - r(t) \la z^2 M_1 \ra_s
 + {1\over 3}\left( 4 \la z R_0 \ra_u- \la z N_0\ra_u \right) \right. \no
&-&\left.{1 \over 8} w(t) \la z \tilde N_1 \ra_u
- {1 \over 4} v(t) \la z^2 \tilde N_1 \ra_u 
+ {1\over 8} r^2(t) \la z^3 \tilde N_1 \ra_u \right\} \no  
\hat R_0(t) &=&\la M_0 \ra_s -{1\over 2} y(t) \la M_1 \ra_s
+{1\over 2}r(t) \la z M_1 \ra_s + {1 \over 3} \left( \la R_0 \ra_u +2 \la
  N_0\ra_u \right)  \co \no 
&+&{1 \over 4} w(t) \la \tilde N_1 \ra_u +{1\over 2} v(t)  \la z \tilde N_1
\ra_u \!-\!{1\over 4}r^2(t) \la z^2 \tilde N_1
\ra_u \; \; , 
\eea
where
\bea
\tilde N_1(t) \!\!&\!\!=\!\!&\!\! N_1(t)/t \;  , \; \; \Sigma = M_K^2+M_\pi^2
\; , \; \; \Sigma_1 = \Sigma+M_\pi^2 \; ,  \; \;
y(t)=\Sigma_1-3t-{M_\pi^2 \Delta \over t} \co \no 
\rho_{\pi K}(t) \!\!&\!\!=\!\!&\!\! \sqrt{\left(1-(M_K\!+\!M_\pi)^2/t \right)
\left(1-(M_K\!-\!M_\pi)^2/t \right) } \co \; \;
r(t)=(t-M_\pi^2)\rho_{\pi K}(t) \co  \no
v(t) \!\!&\!\!=\!\!&\!\! \left(t-{M_\pi^2
    \Delta \over t}\right) r(t) \co \; \; \; \;
w(t)= 3t^2-4t\Sigma_1+5 M_\pi^4 + \Sigma^2 -{M_\pi^4 \Delta^2 \over t^2} \fs
\eea
The brackets $\la \ra$ indicate angular averages defined as
\bea
\la z^n X \ra(s) \!\!&\!\!=\!\!&\!\! {1 \over 2} \int_{-1}^{1} dz
z^n X(\Sigma_1/2-s/2+2 \pp \qq z) \no
\la z^n X \ra_v(t) \!\!&\!\!=\!\!&\!\! {1 \over 2} \int_{-1}^{1} dz
z^n X(v(t,z)) \; \; \; \; \; \; v=s,u \; \; ,
\eea
where
\be
{\pp}^2 = {s\over 4}-M_\pi^2 , \;
{\qq}^2={s\over 4}\left(1-{M_K^2\over s}\right)^2 ; \;
s+u =\Sigma_1 - t ,\; s-u= {M_\pi^2 \Delta \over t} +r(t) z \; .
\label{eq:tch_su}
\ee 
In the definition of the hat functions the function $M_1$ appears. This
is analogous to $M_0$, in the case of the $I=1$ $P$-wave in the $s$
channel, and is necessary to describe the process (\ref{eq:Adef}) in full
generality, for all channels (including the $t \leftrightarrow u$ odd,
$I=1$ $s$--channel). It does not contribute directly to the physical decay
process: its indirect (and small) contribution via the angular average in
the dispersion relation is a consequence of crossing symmetry.

\vskip 0.5cm
\noindent {\bf 3.} If one is only interested in the low--energy region,
neglecting the inelastic channels is a good approximation: then the
solution of the dispersion relation for each of the functions is well
approximated by the Omn\`es function times a polynomial \cite{omnes}. It is
therefore convenient to write the dispersion relation for the functions
divided by the corresponding Omn\`es function (see \cite{eta3pi} for a
detailed discussion of this point, although in a different framework), in
the following form:
\bea
M_0(s) &=& \Omega_0^0(s,s_0)\left\{a+b(s\!-\!s_0)+{(s\!-\!s_0)^2 \over
    \pi} \int_{4M_\pi^2}^{\Lambda_1^2} {\sin \delta_0^0(s') \hat M_0(s') ds'
    \over |\Omega_0^0(s',s_0)| (s'\!-\!s)(s'\!-\!s_0)^2} \right\} \no
N_0(s) &=& \Omega_0^{1/2}(s)\left\{ {s^2 \over
    \pi} \int_{(M_K+M_\pi)^2}^{\Lambda_2^2} {\sin \delta_0^{1/2}(s') \hat
    N_0(s') ds' \over |\Omega_0^{1/2}(s')| (s'-s)s'^2} \right\} \no
N_1(s) &=& \Omega_1^{1/2}(s)\left\{ {s \over
    \pi} \int_{(M_K+M_\pi)^2}^{\Lambda_2^2} {\sin \delta_1^{1/2}(s') \hat
    N_1(s') ds' \over |\Omega_1^{1/2}(s')| (s'-s)s'} \right\} \no
R_0(s) &=& \Omega_0^{3/2}(s)\left\{ {s^2 \over
    \pi} \int_{(M_K+M_\pi)^2}^{\Lambda_2^2} {\sin \delta_0^{3/2}(s') \hat
    R_0(s') ds' \over |\Omega_0^{3/2}(s')| (s'-s)s'^2} \right\} \; \; .
\label{eq:disprel}
\eea
$\Omega^I_\ell(s)$ is the Omn\`es function \cite{omnes}, defined as
\bea 
\Omega^0_0(s,s_0) &=& \exp\left\{{(s-s_0) \over \pi}
\int_{4M_\pi^2}^{\tilde\Lambda_1^2} ds' {\delta^0_0(s') \over (s'-s_0)
(s'-s)} \right\} \; \; \no 
\Omega^I_\ell(s) &=& \exp\left\{{s\over \pi}
\int_{(M_K+M_\pi)^2}^{\tilde\Lambda_2^2} ds' {\delta^I_\ell(s') \over
s'(s'-s)} \right\} ~~~~~~ ~~~~~~ ~~ I={1\over2},{3\over2} \fs 
\eea 

All functions are subtracted at $s=0$ with the only exception of $M_0$,
where the subtraction point $s_0$ is left unspecified. In the
following we use $s_0=M_\pi^2$. The fact that only $M_0$ depends on
subtraction constants does not have any deep reason: the splitting of
polynomial terms of $T^+$ between the various functions $M_0$, $N_{0,1}$
and $R_0$ is arbitrary, and we have merely used this freedom to remove them
from the latter three. The final result does not depend on this choice
\cite{eta3pi}. All the dispersive integrals above have been cut off at
energies $\Lambda_{1,2}$ and $\tilde \Lambda_{1,2}$ -- numerical
values will be given below.

\vskip 0.5cm
\noindent {\bf 4.} If the $\pi \pi$ and $K\pi$ phase shifts, the cutoffs
$\Lambda_{1,2}$, and the subtraction constants $a$ and $b$ are given, the
dispersion relations (\ref{eq:disprel}) can be solved numerically. Such a
solution gives the amplitude $T^+(s,t,u)$ at any point (provided it is far
enough from the inelastic thresholds) of the Mandelstam plane, in
particular at the physical point. The crucial new inputs here are the two
subtraction constants: the phase shifts are known with sufficient accuracy,
whereas the choice of the cutoffs is dictated by the inelastic
thresholds. Before proceeding, we have to discuss how these two subtraction
constants can be determined. If they could be calculated with better
accuracy than the physical amplitude itself, then this would represent for
our method a clear advantage.

For one of the two subtraction constants this is the case.
The soft--pion theorem relates the amplitude at the SPP to the
$K \to \pi$ amplitude up to terms of order $M_\pi^2$. We can therefore
write
\be
-{1\over 2 F_\pi} {\cal A}(K \to \pi) = a + {1\over 3}
\left[N_0(M_K^2)+2R_0(M_K^2)\right] +{\cal O}(M_\pi^2) \co
\label{eq:SPP}
\ee
which shows that $a$ is indeed directly related to a quantity that is
calculable (more easily than the decay amplitude itself), e.g. on the
lattice. The relation (\ref{eq:SPP}) illustrates the strength of the
soft--pion theorem: although the process involves a kaon, the relation
is based on the use of the $SU(2)$ symmetry, and therefore suffers
from corrections of order $M_\pi^2$ only.

The key of the problem is how to calculate $b$. This constant is related
to the derivative in $s$ of the amplitude $T^+$ at the SPP. 
The calculation of $b$ requires the evaluation of the physical amplitude
$T^+$ at an unphysical point, via analytic continuation. While this is easy
to do with an analytical method like CHPT, it is practically impossible
with a numerical method, like the lattice. However there is a Ward identity
that relates this derivative to a Green function that is directly
calculable:
\be
{\partial \over \partial s} T^+(s,\Sigma-s,M_\pi^2)_{|s=M_\pi^2} ={1 \over
  2} C(M_\pi^2,M_K^2,M_\pi^2) +{\cal O}(M_\pi^2) \co
\ee
where $C(s,t,u)$ is an amplitude defined as:
\be
{i \over F_\pi} \int \!\!dx e^{ip_1x} \la \pi(p_2) | T {\cal H}_W^{1/2}(0)
  A^\mu(x) |K(q_1) \ra = ip_1^\mu B +iq_1^\mu C + i q_2^\mu D \; ,
\label{eq:BCD}
\ee 
where $A^\mu(x)$ is the axial current that couples to the pion removed from
the outgoing state. By making the $p_2$ momentum soft one can also derive a
soft-pion theorem which relates the four-point function in
Eq. (\ref{eq:BCD}) to a three-point function. Unfortunately the function
$C$ cannot be singled out from this relation.

We are not aware of any attempts to calculate $b$.  
In order to illustrate our method we proceed by fixing $b$ at a certain
value and then varying it within a fairly wide range. To fix the central
value and the range we use CHPT as a guide.
At leading order, CHPT dictates the following relation between $a$ and $b$:
\be
b= {3 a\over M_K^2-M_\pi^2} \left(1+ X + {\cal O}(M_K^4)
\right) \fs 
\label{eq:ab_CHPT}
\ee
The size of the correction is at the moment unknown, but nothing
protects it from being of order $M_K^2$: $X=M_K^2/(16 \pi^2 F_K^2) x$,
with $x$ expected to be of order one. An explicit 
calculation in CHPT yields\footnote{We have dropped the contribution coming
  from the weak mass term -- more on this below.} \cite{FF}:
\bea
x\!&\!=\!&\!\frac{383}{9} -
  \frac{8021}{54} \ln {4\over3} +\frac{1}{3} \ln {M_K^2\over M_\pi^2} 
- \frac{8}{3} ( \bar N_5 +2 \bar N_7- \bar N_9
-4  \bar N_{10}-4  \bar N_{11}) \no
&+&2( \bar N_{19} - \bar N_{20}) 
- \frac{4}{3} (2 \bar N_{21} + \bar N_{22} +2 \bar N_{23}) +
{\cal O}\left({M_\pi^2 \over M_K^2}\right) \; , 
\label{X1loop}
\eea
where $\bar N_i=16\pi^2 N_i^r(M_K)$ are the renormalized low-energy
constants introduced in \cite{EKW}.
Since we lack information on many of the constants, the CHPT calculation
(\ref{X1loop}) does not allow us to do more than an order of magnitude
estimate for $b$.

\vskip 0.5cm
\noindent {\bf 5.}
\begin{figure}[t] 
\leavevmode \begin{center}
\includegraphics[width=12cm]{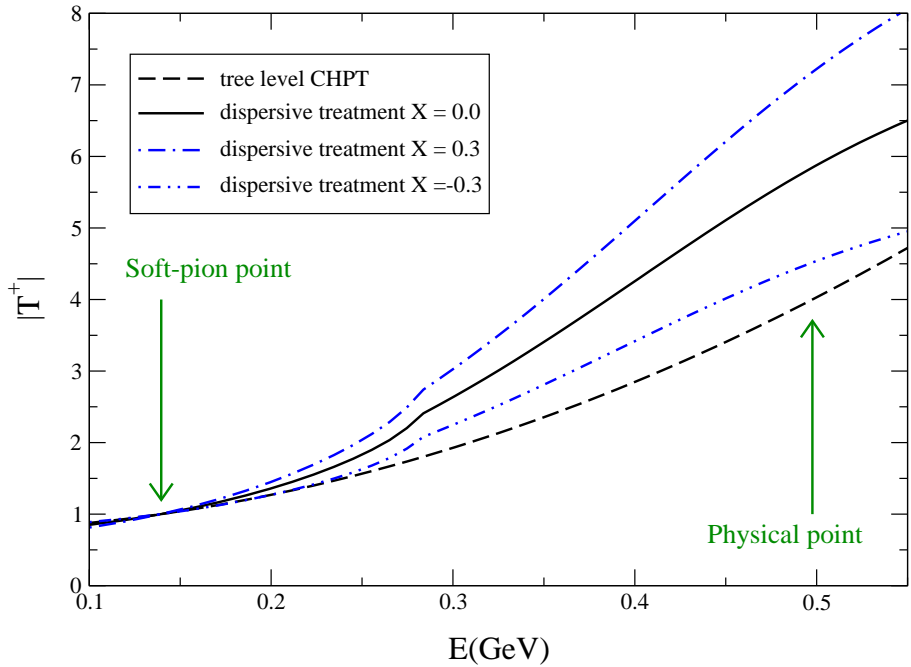}
\caption{\label{fig1}The function $|T^+(s,t,u)|$ plotted {\em vs.} $E=\sqrt{s}$
  along the line of constant $u=M_\pi^2$: the result of our numerical study
  for different values of $X$ are compared to tree level CHPT.} 
\end{center}
\end{figure}
In our numerical study we have used $ X = \pm 30\%$. The normalization of the
amplitude is irrelevant here, and we have fixed it to
$T^+(M_\pi^2,M_K^2,M_\pi^2)=1$. The $\pi \pi$ phase shifts are taken from
\cite{ACGL}, with the scattering lengths determined in \cite{CGL}, and the
$\pi K$ phase shifts from \cite{JOP}. For the cutoffs we have used
$\Lambda_1=1$ GeV, $\Lambda_2=1.3$ GeV, and $\tilde \Lambda_i=1.05
\Lambda_i$. Our results are shown in Fig. \ref{fig1}, where we have plotted
$|T^+(s,\Sigma-s,M_\pi^2)|$ versus $s$, comparing our numerical solution of
the dispersion relations to the CHPT leading order formula.  The latter is
what has been used so far whenever a number for the $K \to \pi \pi$ matrix
element extracted from the lattice has been given. Our treatment shows that
large corrections with respect to leading--order CHPT are to be
expected. One source of large corrections is the Omn\`es factor due to $\pi
\pi$ rescattering in the final state \cite{PP,truong}. The other
potentially dangerous source is represented by $X$, the next-to-leading
order correction to the relation (\ref{eq:ab_CHPT}) between $a$ and
$b$. The latter could (depending on the sign) in principle
double, or to a large extent reabsorb the correction due to final
state interactions. The dependence on $X$ is well described by the
following linear formula: 
\be 
{|{\cal A}(K \to \pi \pi)|
\over |{\cal A}^{\mbox{\tiny LO CHPT}}(K \to \pi \pi)|} = 1.5\left(1+0.76
X\right) \co
\label{eq:linear}
\ee 
after having normalized both amplitudes to $T^+(M_\pi^2,M_K^2,M_\pi^2)=1$.
The evaluation of the uncertainties to be attached to the numbers in
Eq. (\ref{eq:linear}) is in progress \cite{wip}. At the moment, however,
the main source of uncertainty is the fact that $X$ is largely unknown.

One of the outcomes of the present analysis is that the effects
embodied in the functions $N_{0,1}$ and $R_0$ have turned out to be
very small: if we drop them altogether, the numbers in
Eq. (\ref{eq:linear}) change from 1.5 to 1.4 and from 0.76 to 0.75. 
Notice that these effects are in principle of order $M_K^2$, as can be seen
in Eq. (\ref{eq:SPP}), and that they are not a priori negligible.  On the 
other hand this result is very much welcome, because the size of these
functions depends both on the $\pi K$ phases (which are less well known
than the $\pi \pi$ ones) and on the choice of the cutoff $\Lambda_2$, which
may induce large uncertainties.

\vskip 0.5cm
\noindent {\bf 6.} The authors of \cite{Bernard} have given a recipe to
remove by hand the contribution of the weak mass term to the $K \to \pi$
transition. This step is necessary since at tree level the weak mass term
does not contribute to the $K \to \pi \pi$ amplitude. The procedure
suggested in \cite{Bernard} involved the calculation of the $K \to$ vacuum
transition. In our framework this procedure is unnecessary. To show this it
is useful to work with the tree-level CHPT amplitude (weak mass
contribution only, see \cite{EKW} for the notation):
\be 
T^+_{c_5}(s,t,u) = \frac{-ic_5 \Delta}{F_\pi^2}
\left[\frac{s}{q_2^2-M_K^2}+1\right] \co
\label{c5tree}
\ee
where we have restored $q_2^2\neq 0$ to show explicitly the presence of the
kaon pole. 
At the physical point this amplitude vanishes. On the other hand, if we
calculate this amplitude at the SPP, from the
corresponding $K \to \pi$ amplitude, we cannot get the kaon-pole term, and
therefore would obtain a nonvanishing contribution at the physical
point. Hence the cure proposed in \cite{Bernard}. The fact that
the $K \to \pi$ amplitude does not contain the pole term does not
contradict the soft-pion theorem: at the SPP, for $q_2^2=0$, the pole term
is of order $M_\pi^2$, which is beyond the accuracy the theorem
guarantees. 

The framework proposed here relies on the presence of two subtraction
constants, that we have related to the amplitude and its derivative at the
SPP. If one determines {\em both} constants at tree level in CHPT
(\ref{c5tree}), one finds that the subtraction polynomial vanishes at the
physical point, without further ado. This cancellation does not happen if
the subtraction constants are given to one loop -- but it does not need to:
as is well known \cite{BPP}, the $K \to \pi \pi$ amplitude does contain
contributions proportional to $c_5$ at one loop.

\vskip 0.5cm
\noindent {\bf 7.} 
In the present paper we have set up a dispersive framework for the
$K\rightarrow \pi\pi$ amplitude that allows one to evolve the amplitude
from the soft-pion point (where it is given by the $K \rightarrow \pi$
amplitude) to the physical point, taking into account all the main physical
effects. As we have pointed out, this evolution is on safe ground only if a
second input is made available: the derivative of the amplitude at the
soft-pion point which, to the best of our knowledge, has not been
calculated so far. We have calculated this second subtraction constant to
next-to-leading order in CHPT. Given the presence of unknown low-energy
constants, we cannot use this expression for more than an order of
magnitude estimate. Our numerical work, however, shows that the amplitude
at the physical point depends strongly on the value of the slope at the
SPP, see Fig. 1 and related discussion in the text. A nonperturbative
calculation of the second subtraction constant $b$ is necessary in order to
obtain an accurate result with this method. We have provided a Ward
identity which might be useful in this respect.

Lattice calculations of the $K\rightarrow \pi\pi$ amplitude made so far
\cite{lattice} rely on tree-level CHPT to relate the calculated
$K\rightarrow \pi$ matrix elements to the physical decay amplitude. The
method proposed here improves this scheme by combining input from the
lattice with dispersive techniques, thereby providing a fully consistent
treatment of final state interactions in $K\rightarrow \pi\pi$. Given the
two subtraction constants, the dispersion relations can be solved
numerically to good accuracy. Recently, a direct calculation of the $K \to
\pi \pi$ matrix element on the lattice has been proposed in
Ref. \cite{lellu} -- this method does not rely on CHPT. Other lattice
methods, which also do not rely on the evaluation of the $K \to \pi$
amplitude had also been proposed previously \cite{rome}. Each of these
methods presents different technical problems in its practical
implementation \cite{golterman}, and it is difficult to predict which one
will lead to a reliable calculation of the $K \to \pi \pi$ amplitude. We
hope that the present work will stimulate further efforts to calculate the
subtraction constants $a$ and $b$, either on the lattice, or by other
nonperturbative methods.

\vskip 0.5cm \noindent 
{\bf Acknowledgements}~~ It is a pleasure to thank
J.~Gasser, G.~Isidori, E.~Pallante, T.~Pich, and D.~Wyler for interesting
discussions. This work was supported by the Schweizerische Nationalfonds,
by TMR, BBW-Contract No. 97.0131 and EC-Contract No. ERBFMRX-CT980169
(EURODA$\Phi$NE).

\end{document}